# MAAS-SFRThelper: An Integrated ESAPI Plugin for Structure Generation, Optimization, and Evaluation of Spatially Fractionated Radiation Therapy

Japan K. Patel[1], Anthony Magliari[2], Gregory Gill[3], Todd A. Wareing[2], Tenzin Kunkyab[4], Caleb Raman[5], Ilias Sachpazidis[6], Peter Szentivanyi[2], Ryan Clark[2], Pierre Lansonneur[2], Arjun Karnwal[7], Michael Kudla[8], Sergejs Unterkirhers[9], Junqi Song[10], Jun Yang[10], and Matthew C. Schmidt[1,11]

[1]Gateway Scripts, MO, USA
[2]Medical Affairs, Varian Medical Systems, CA, USA
[3]Department of Radiation Oncology, Intermountain Health, UT, USA
[4]Department of Radiation Oncology, Icahn School of Medicine at Mt. Sinai, NY, USA
[5]Biological Sciences Division, University of Chicago, IL, USA
[6]Department of Radiation Oncology, University of Freiburg, Germany
[7]Department of Biomedical Engineering, University of Southern California, CA, USA
[8]Radiation Therapy Program, BC Cancer, Canada
[9]Clinic Hirslanden, Switzerland
[10]Foshan Fosun Chancheng Hospital, Foshan City, China
[11]School of Medicine, Washington University in St. Louis, MO, USA

## Abstract

**Purpose:** Spatially fractionated radiation therapy (SFRT) planning requires three coordinated tasks: generation of high-dose sphere structures, position-aware optimization, and peak-valley dose ratio evaluation. In practice, clinicians and researchers address these tasks with a combination of closed-source scripts, research codes, and manual calculation. The absence of a unified, commercially deployable toolkit remains a barrier to multi-institutional SFRT trials. We present MAAS-SFRThelper, a shared-source Eclipse Scripting Application Programming Interface (ESAPI) plugin that integrates structure generation, geometric-aware optimization, and peak-valley dose ratio evaluation for SFRT into a single workflow inside Varian's Eclipse treatment planning system.

**Acquisition and Validation Methods:** MAAS-SFRThelper is a Windows Presentation Foundation (WPF) application built on the Model-View-ViewModel (MVVM) pattern, implemented in C# against ESAPI for Eclipse 15.6 and later. The plugin currently contains five task-oriented tabs that share common services for sphere extraction and objective creation. The SphereLattice tab generates sphere lattices using five placement patterns: hexagonal close-packed, simple cubic, alternating cubic, centroidal Voronoi tessellation, and a constraint-based Monte Carlo method. The SCART tab creates contracted spindle-like boost volumes for stereotactic central ablative radiation therapy. The Optimization tab auto-populates structure list, searches over candidate lattice positions using a four-metric geometric surrogate score, and triggers Eclipse VMAT optimization along with dose calculation on the best candidate. The Evaluation tab implements four analysis modes including 1D, 2D, and 3D PVDR analysis. We validated all functionality on digital phantoms against analytic ground truth. In phantom testing, the repositioned lattice achieved 30% higher PVDR and 22% lower OAR mean dose.

**Data Format and Usage:** The plugin is distributed as source code that compiles to a single dynamic-link library (DLL); all third-party dependencies are bundled at build time. At first launch, users review the Varian Limited Use Software License Agreement and enter an access code that is emailed upon request. The plugin accepts the patient structure set and, where applicable, a plan with calculated dose; outputs include new structures written directly to the structure set through ESAPI and summary statistics exported as comma-separated value (CSV) files. Source code and documentation are publicly available on GitHub under the Varian Limited Use Software License Agreement (LUSLA).

**Potential Applications:** MAAS-SFRThelper supports clinical lattice SFRT planning with consistent workflows across institutions and standardized peak-valley dose ratio reporting for multi-institutional trials. The shared-services architecture enables community contributions of new placement patterns, evaluation metrics, and validation datasets. These features represent a practical step toward the dosimetric consistency called for in the 2024 NRG Oncology/AAPM consensus on SFRT. The plugin also serves as a platform for research extensions to GRID therapy, minibeam radiation therapy, and other heterogeneous dose-delivery modalities.

## 1. Introduction

Bulky, radioresistant, and recurrent tumors remain a persistent clinical challenge for conventional radiation therapy [1]. Soft-tissue sarcomas [2], unresectable abdominal and pelvic masses [3], and recurrences in previously irradiated sites [4] often do not respond adequately to uniform dose coverage of the target planning volume. Spatially fractionated radiation therapy (SFRT) intentionally creates a heterogeneous dose distribution with regions of high dose (peaks) interspersed with regions of low dose (valleys) [5]. The technique originated as GRID therapy, delivered with patterned brass blocks [6], and is now implemented using volumetric-modulated arc therapy (VMAT) to produce three-dimensional lattice patterns within the target [7]. Minibeam radiation therapy extends the same principle to a finer spatial scale using sub-millimeter planar beams [1]. Clinical interest in SFRT has grown in recent years, driven by encouraging outcomes in challenging cases [8] and by biological evidence that heterogeneous dose distributions can elicit bystander [9] and abscopal responses [10] beyond the direct effects of the delivered dose. Translating SFRT into routine clinical practice is not straightforward.

Clinical evidence supporting SFRT has accumulated steadily over the past decade. A systematic review by Iori et al. [3] pooled 81 patients across 12 reports and concluded that lattice radiation therapy is safe, with a median lesion reduction of approximately 50% or greater at 3-6 months when a complete response was not achieved. A subsequent meta-analysis of seven single-arm studies covering 187 patients reported pooled complete response rates of 42% for lattice therapy followed by external beam radiation and 23% for fractionated lattice therapy alone, with a grade 3-4 adverse event rate of 3.4% [11]. Single-center cohorts continue to report favorable outcomes; the Mayo Clinic series of Owen et al. [8] covered 176 patients and 186 treated sites with sustained tumor shrinkage and acceptable toxicity. A 2024 NRG Oncology/AAPM consensus identified SFRT as a technique for which prospective multi-institutional trials are needed [5]. Several trials are now active, including a Phase 2 study at the University of Cincinnati [12]. The evidence base is still developing, but the direction is consistent: SFRT produces meaningful tumor responses in the bulky and radioresistant settings where conventional radiation therapy has limited options.

The idea of delivering spatially heterogeneous dose distributions dates to Köhler in 1909 [13]. The modern re-emergence came with Mohiuddin et al. [6], who demonstrated high-dose spatially fractionated radiation using patterned brass blocks as a single-fraction boost. Wu et al. [7,14] extended the concept into three dimensions by placing discrete high-dose spheres inside the gross tumor volume and delivering them with multileaf collimator (MLC)-based techniques, giving rise to lattice radiation therapy. Clinical SFRT today is most often delivered as photon-based GRID or lattice therapy using VMAT [1,15]; proton-based [16,17] and heavy-ion [18] implementations have also been reported. Common to all approaches is the deliberate creation of dose peaks interleaved with lower-dose valleys, quantified through the peak-to-valley dose ratio (PVDR) [1].

The growing clinical adoption of SFRT has motivated development of software tools that address parts of the planning workflow. Most progress has been on automated sphere placement. Protocol-specific scripts now generate lattices in minutes rather than the 20-30 minutes required by manual placement [19,20], and several groups have developed tools that personalize lattice geometry to tumor size and shape - iterating over vertex size and spacing [21,22], maximizing vertex count while minimizing OAR overlap [23], enforcing geometric and dosimetric constraints during sampling [24], applying biased Monte Carlo searches with user-selectable optimization goals [25], or standardizing vertex dimensions for dense fractionated delivery [26]. On the optimization side, Zhang et al. [27] modeled vertex positions as differentiable sigmoid functions and jointly optimized them with plan variables to improve PVDR and OAR sparing, with a related approach developed for proton pencil-beam scanning through total-variation and L1 dose regularization [28]. Commercial automated planning engines have also been evaluated for lattice SFRT [29]. Evaluation of SFRT plans centers on the peak-to-valley dose ratio, though its computation is not standardized: groups report PVDR from hand-selected dose profiles, dose thresholds applied to the full three-dimensional distribution, or structure-based averages over vertex and valley contours [1,3]. A 2024 NRG Oncology/AAPM consensus identified this lack of standardized dosimetric reporting as a barrier to prospective multi-institutional trials [5]. The field has made substantive progress on individual parts of the planning workflow, with each tool addressing a specific clinical need or protocol.

Our goal was to build a single, freely available package that covers structure generation, optimization, and evaluation inside a commercial treatment planning system with multiple pattern options that accommodate different institutional preferences, integrated peak-valley-aware optimization, and a consistent evaluation pipeline that computes PVDR and related metrics the same way every time. MAAS-SFRThelper is organized around three core capabilities. The SphereLattice tab supports five placement patterns: hexagonal close-packed, simple cubic, alternating cubic [19], centroidal Voronoi tessellation in three dimensions (CVT), and an experimental constraint-based Monte Carlo placement [24]. The CVT [30] implementation adapts an open-source two-dimensional generator to three dimensions [31]. The SphereLattice tab also generates matching void structures for valley dose control, and two additional tabs cover stereotactic central ablative radiation therapy (SCART) [32] and experimental rod-based patterns. The Optimization tab introduces geometric surrogate metrics that characterize a candidate lattice configuration without requiring dose calculation; it performs a grid search over sphere positions and applies the best configuration for VMAT optimization and dose calculation inside the treatment planning system. The Evaluation tab provides a geometrically grounded three-dimensional peak-valley classification with four analysis modes, computes volume-weighted effective PVDR, heterogeneity index, coefficient of variation, and mean peak separation, and supports an onion-layer shell analysis for radial dose heterogeneity. Early versions of the toolkit focused on sphere generation [33]; subsequent extensions added the evaluation tab [34] and presented the optimization tab with full methodology [35]. The toolkit has also been used in external work, including a recent dosimetric verification study with polymer gel phantoms [36]. The compiled software, complete source code, and documentation are publicly available at https://github.com/Varian-MedicalAffairsAppliedSolutions/MAAS-SFRThelper under the Varian Limited Use Software License Agreement (LUSLA).

The remainder of this paper is organized as follows. Section 2 describes the software architecture and structure generation, optimization, and evaluation tabs, followed by a phantom-based validation of select features. Section 3 covers installation, system requirements, data formats, and usage notes. Section 4 discusses the scope and limitations of the current implementation and outlines planned extensions. Section 5 summarizes and points to future work.

## 2. Acquisition and Validation Methods

This section describes the software architecture of MAAS-SFRThelper and its structure generation, optimization, and evaluation tabs. We first outline the software stack and the shared services that underpin them (Section 2.1). We then describe the structure-generating tabs (Section 2.2), the optimization tab (Section 2.3), and the evaluation tab (Section 2.4). Each subsection covers the algorithms, the ESAPI interactions, the outputs produced, and phantom-based validation of select features.

### 2.1 Software Architecture

MAAS-SFRThelper is a Windows Presentation Foundation (WPF) plugin written in C# that targets .NET Framework 4.8 and the Eclipse Scripting Application Programming Interface (ESAPI) for Eclipse 15.6 or later. The code follows the Model-View-ViewModel (MVVM) pattern using the Prism framework [37] and draws on Helix Toolkit [38] for three-dimensional visualization of structures and dose, and NLog [39] for application logging. The plugin compiles to a single DLL; all third-party dependencies are bundled at build time.

MAAS-SFRThelper presents a tabbed interface in which each tab corresponds to a distinct SFRT task as shown in Figure 1. The tabs are ordered to reflect the primary clinical workflow. SphereLattice places lattice spheres and optional void structures. The Optimization tab repositions the lattice through a geometric surrogate search, runs VMAT optimization, and performs dose calculation. The Evaluation tab characterizes the resulting dose distribution with PVDR and heterogeneity metrics. Two additional tabs support alternative workflows. SCART contracts the gross tumor volume (GTV) to form the stereotactic central ablative RT volume or Spindle Target Volume (STV) and runs its own VMAT optimization and dose calculation internally [32]. RapidRods produces experimental rod-based structures to simulate brachytherapy.

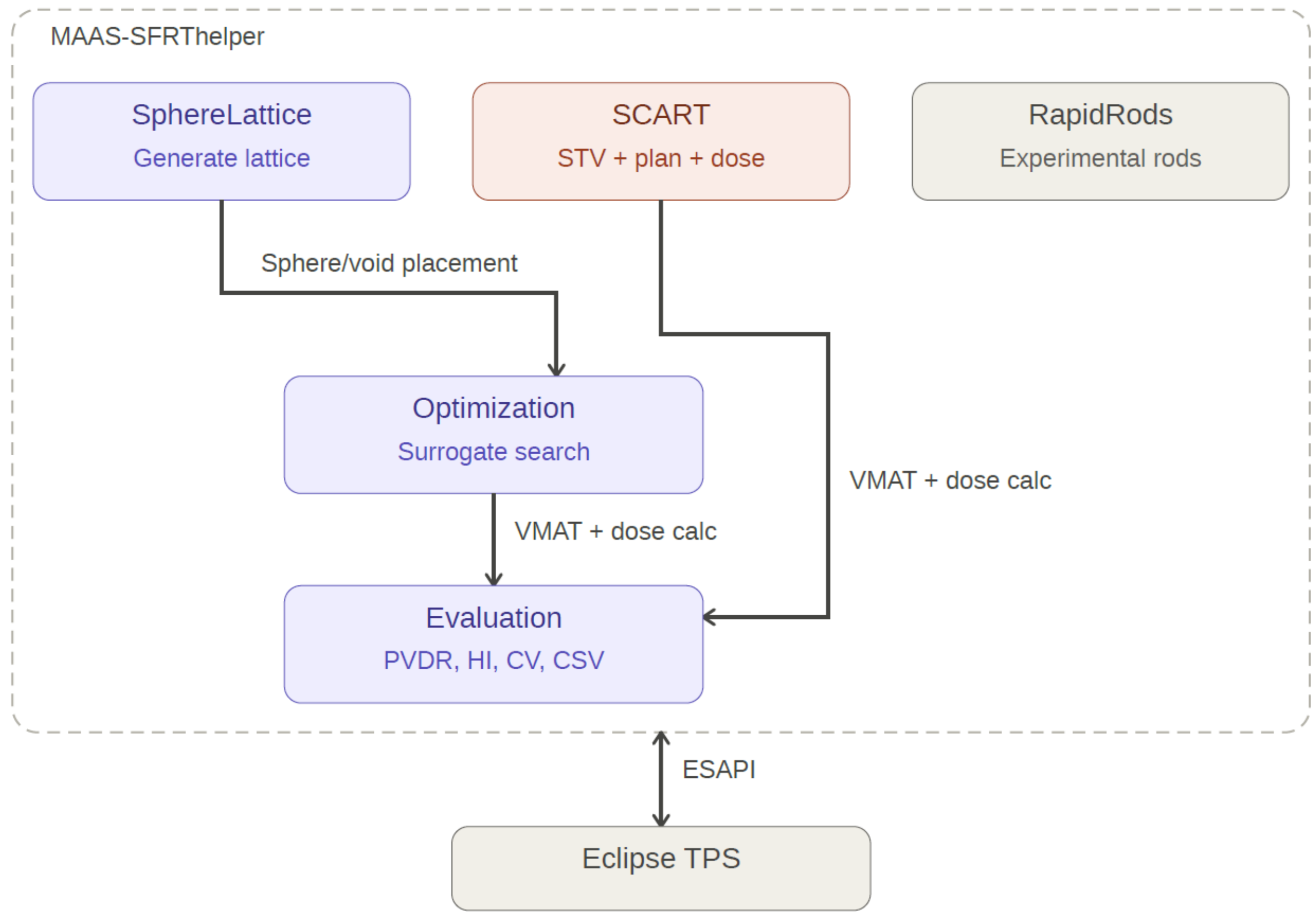

Figure 1: MAAS-SFRThelper plugin architecture and workflow.

For the standard lattice SFRT workflow, the user proceeds through SphereLattice, Optimization, and Evaluation in sequence: a lattice structure produced in SphereLattice flows into Optimization for repositioning, and the resulting plan and dose flow into Evaluation for analysis. SCART bypasses Optimization and hands its completed plan directly to Evaluation. Throughout, Eclipse serves as the shared data layer where structures, images, and dose are read and written through ESAPI, and the outputs of each tab are written back to Eclipse through the same interface.

Eclipse enforces single-thread affinity on ESAPI calls: every interaction with the patient model must execute on the dispatcher thread that originally received the ScriptContext. To keep the interface responsive while honoring that constraint, we built a lightweight dispatcher wrapper, EsapiWorker, that schedules ESAPI work either asynchronously (Run) or with a blocking wait (RunWithWait), depending on whether the next user action depends on the result. All ESAPI reads and writes in the plugin go through this wrapper. The remaining shared services are domain-specific: SphereExtractor discovers sphere centers and a mean sphere radius from a lattice structure (used by both the Optimization and Evaluation tabs); GeometricSurrogateCalculator computes geometric surrogate scores for candidate lattice placements; OptimizationObjectiveCreator identifies available lattice, valley, and target structures in the structure set for the Optimization tab; and the MayoLattice sub-services implement constraint-based Monte Carlo lattice placement [24].

## 2.2 Structure Generation

The structure generation tabs produce the structures that the optimizer uses to create a peak-and-valley dose pattern in an SFRT plan. SphereLattice tab generates structures using the selected placement pattern, SCART contracts the GTV to produce the STV, and RapidRods produces experimental rod-based structures. The lattice patterns share a common set of inputs: a target structure, a sphere radius, and a partial-sphere acceptance threshold that controls how much each candidate sphere must lie inside the target. The three analytic patterns additionally expose a center-to-center spacing, optional XY grid shifts, and a lateral scaling factor to enforce additional lateral spacing between spheres to improve peak to valley ratios for coplanar delivery. For all grid-based methods, the lattice origin is shifted to the centroid of the target structure so that the lattice center and the target center coincide. Every SphereLattice pattern can write each sphere as a separate structure alongside a combined parent lattice structure and exports vertex coordinates and generation parameters to a CSV file. ESAPI limitations often require the user to forego separate structures due to the structure limit (99) in Eclipse 18.0 and below. Companion void structures allow for valley-dose control during optimization and can be generated for every grid-based pattern.

**Analytic patterns.** The three analytic patterns place spheres deterministically on a crystallographic lattice centered on the target centroid. SCP uses a uniform Cartesian grid with spacing s along each axis. AC uses the same grid but interleaves spheres and voids: the $(i, j, k)$ cell holds a sphere when $(i + j + k)$ is even and a void otherwise. HCP stacks hexagonal layers in an *ABAB* repeating sequence at the ideal $\frac{c}{a} = \sqrt{\frac{8}{3}}$ ratio, with interstitial void positions between layers. For all three patterns, candidate points are kept only if they lie inside an inwardly retracted copy of the target, with the retraction depth controlled by the partial-sphere acceptance threshold. HCP is set as the default option.

**Centroidal Voronoi tessellation.** The method introduces placement variability through Lloyd's iterative relaxation [40], which redistributes vertices inside the target toward a centroidal Voronoi tessellation [41]. Lansonneur originally proposed this methodology for SFRT vertex placement [30]; the three-dimensional implementation adapts the open-source CVTGenerator library from Sachpazidis [31]. Candidate generators are seeded from an HCP grid at the user's selected spacing inside the retracted target, relaxed toward the centroids of their Voronoi regions until convergence, and filtered to remove any generator whose nearest accepted neighbor lies closer than 2.1 times the sphere radius. The HCP seeding determines the approximate sphere count; relaxation adapts positions to target geometry.

**Constraint-based Monte Carlo.** This placement methodology implements the sphere-sampling approach of Deufel et al. [24]. Candidate sphere centers are accepted only when six geometric constraints are simultaneously satisfied: a minimum distance from the GTV boundary (default 5 mm), a minimum distance from any selected OAR (default 10 mm), a minimum center-to-center spacing between spheres (default 30 mm), and two axial separations (default 20 mm and 80 mm) that together discourage stacking of spheres along shared beam paths. Selecting this method exposes an advanced panel where each constraint can be edited individually. Placement proceeds in three stages: the feasible region satisfying the margin constraints is first constructed, per-slice centers of mass of that region are computed, and spheres are sampled iteratively with a centralization pressure that biases placement toward axial centers. An optional gradient-walk refinement repositions each accepted sphere to a local optimum, and runs are reproducible when the user supplies a random seed. This method is currently experimental and under active validation.

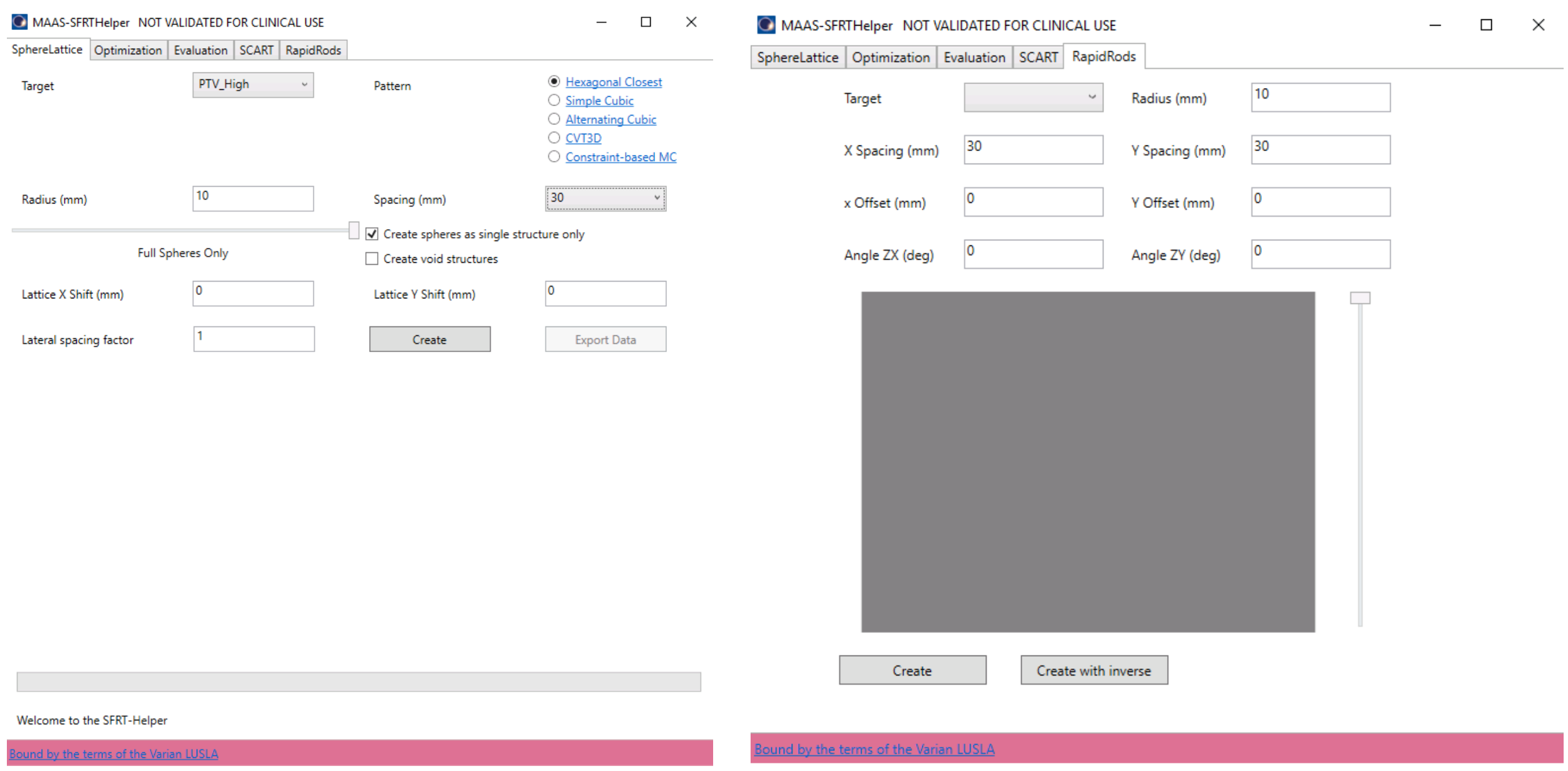

Figure 2a: SphereLattice tab interface.

Figure 2b: RapidRods tab interface.

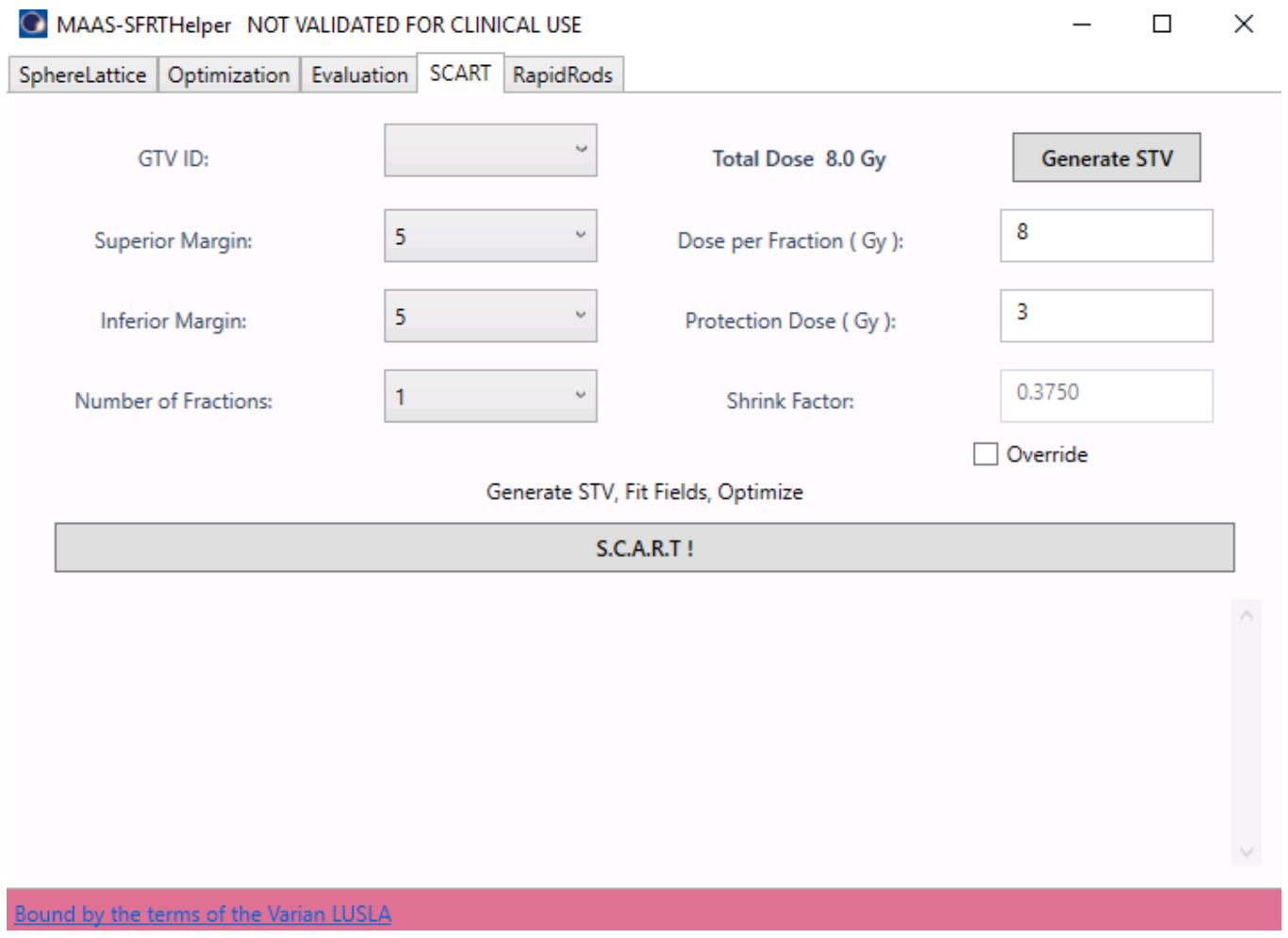

Figure 2c: SCART tab interface.

Figure 2: Interfaces for tabs that generate structures for high dose targeting.

**SCART.** This tab contracts the GTV to produce the STV used as the ablative target volume. The user specifies the GTV, superior and inferior margins in slices, and an ablation-protection dose pair (for example, 21 Gy ablative and 5 Gy protection). Each transverse contour of the GTV is examined in polar coordinates and contracted radially by a factor derived from the protection-to-ablation dose ratio, as described by Song et al. [32]. The tab then builds VMAT beams, runs plan optimization, and performs dose calculation internally, producing a self-contained SCART plan without requiring the Optimization tab.

**RapidRods.** This is an experimental tab that produces cylindrical rod structures in place of spherical vertices. The user draws circular cross-sections on a two-dimensional plane, and each circle is extruded along the superior-inferior axis between user-selected slice bounds. Optional tilt angles about the x- and y-axes apply a per-slice offset to produce angled rods. The final structure is written after cropping from the target. RapidRods is included for exploratory investigation of rod-based SFRT geometries for possible substitution when brachytherapy is not available. This method is not validated or recommended for clinical use.

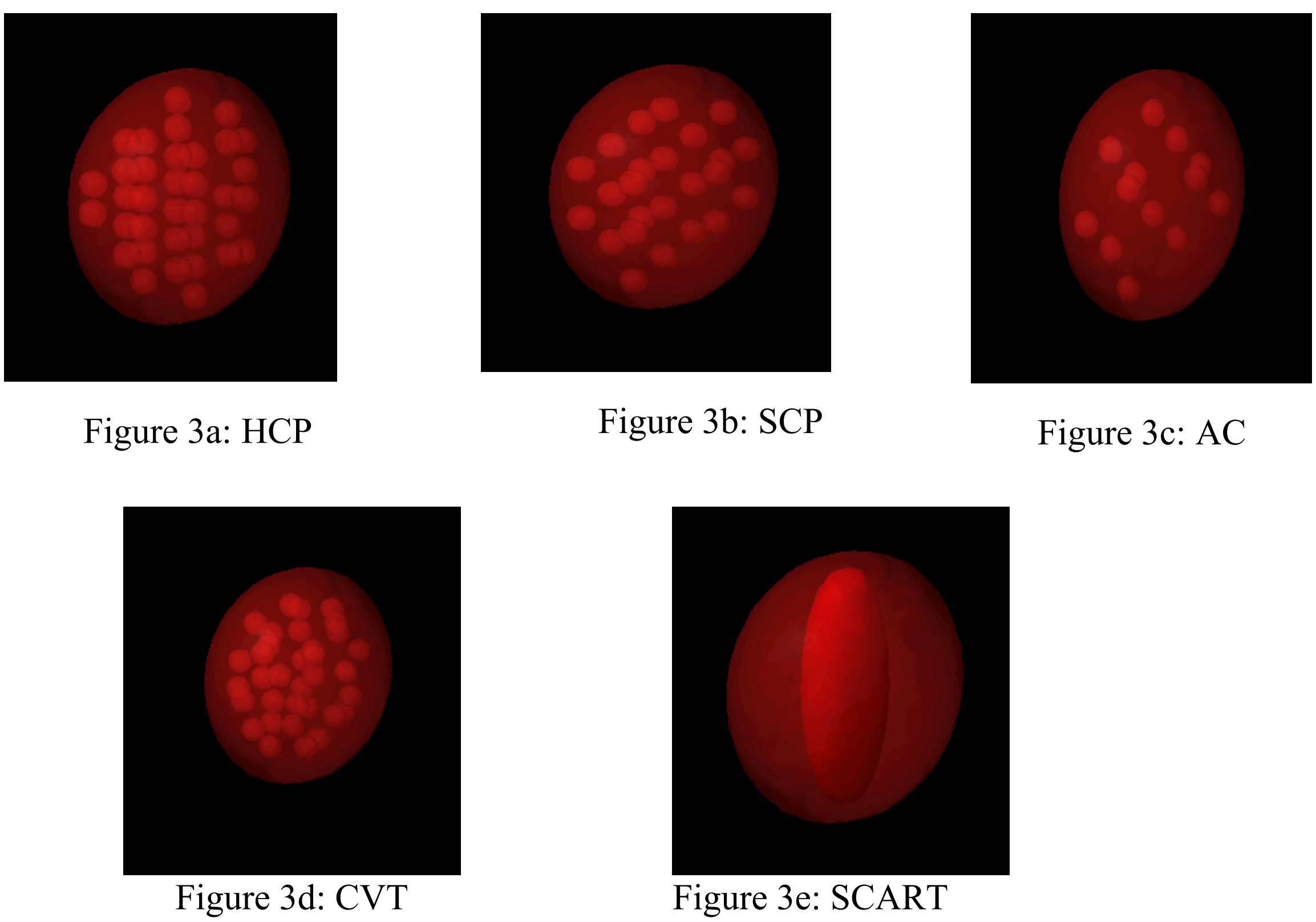

Figure 3a: HCP

Figure 3b: SCP

Figure 3c: AC

Figure 3d: CVT

Figure 3e: SCART

Figure 3: Structure generation patterns (excluding experimental ones).

**Validation.** Structure generation was validated on a phantom consisting of an ellipsoidal target (PTV_High, 1013 cc) centered inside a cylindrical body, with a 15-mm-diameter cylindrical OAR positioned 68 mm from the target surface. All four SphereLattice patterns were run with r = 7.5 mm and 30 mm grid spacing. Table 1 reports sphere count and nearest-neighbor statistics for each pattern. The three analytic patterns (HCP, SCP, AC) produced zero-variance nearest-neighbor distances matching their theoretical values: 30.00 mm for HCP and SCP, and 42.43 mm for AC, confirming exact grid fidelity. HCP placed 35 spheres with inter-layer Z spacing of 24.5 mm, consistent with the ideal $\frac{c}{a} = \sqrt{\frac{8}{3}}$ ratio. CVT preserved the HCP sphere count while Lloyd relaxation shifted positions a mean of 6.64 mm from the HCP seeds; the resulting nearest-neighbor distances ranged from 21.41 to 28.75 mm, and all pairwise separations exceeded the 2.1

× radius minimum-separation filter. For SCART, the STV volume produced at a protection-to-ablation ratio of 3:8 (shrink factor 0.375) was 141.4 cc. The polar contraction scales each transverse contour area by the square of the shrink factor; the predicted volume of 142.5 cc agrees to within 0.8%. RapidRods was excluded from validation as noted above.

| **Pattern** | HCP | SCP | AC | CVT |
|---|---|---|---|---|
| **Spheres** | 35 | 24 | 13 | 35 |

Table 1: Number of spheres per pattern.

**LatticeRT**. A companion web application, LatticeRT, shown in Figure 4, reimplements the four validated placement patterns (HCP, SCP, AC, CVT) in JavaScript with WebGL rendering. It provides a four-viewport display (axial, sagittal, coronal, and interactive three-dimensional) that loads CT and RTSTRUCT DICOM files directly without requiring Eclipse. The application classifies generated spheres into peak, warm, and cold zones and exports generated lattices as derived RTSTRUCT files that preserve the original frame of reference. The tool runs entirely in the browser and is distributed separately under Varian LUSLA. This repository is hosted at https://github.com/Varian-MedicalAffairsAppliedSolutions/LatticeRT.

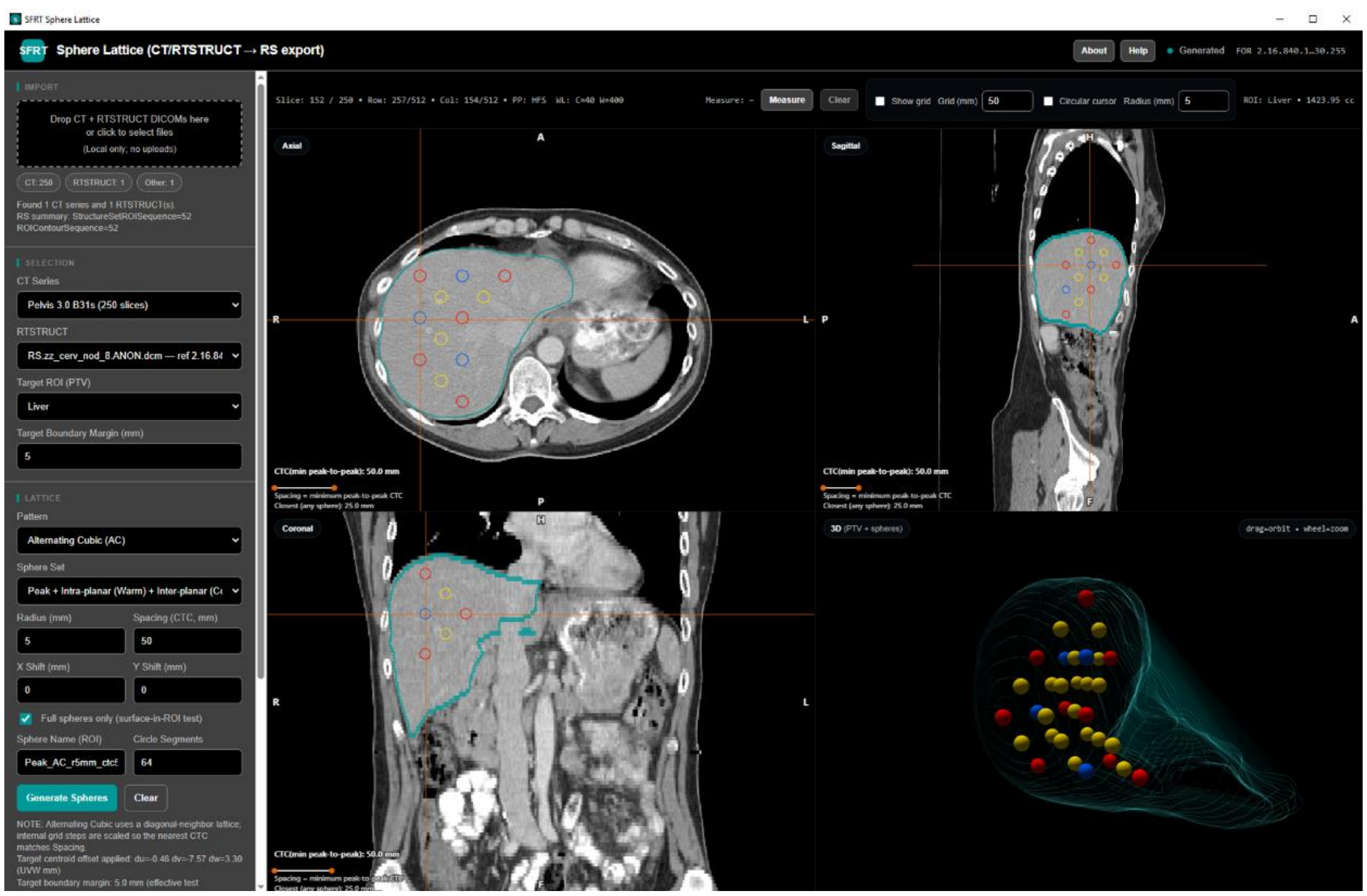

Figure 4: SFRT Sphere Lattice companion JavaScript application.

## 2.3 Optimization

With the lattice structure written into the structure set, the next step is positioning it for optimal plan quality. Zhang et al. [27] demonstrated that PVDR varies substantially with lattice position for the same target. The Optimization tab brings position-aware planning inside Eclipse. It evaluates candidate lattice positions through four geometric surrogate metrics, writes the best configuration back into the structure set, populates an editable objective table for user-specified dose constraints, and invokes VMAT optimization with dose calculation via ESAPI.

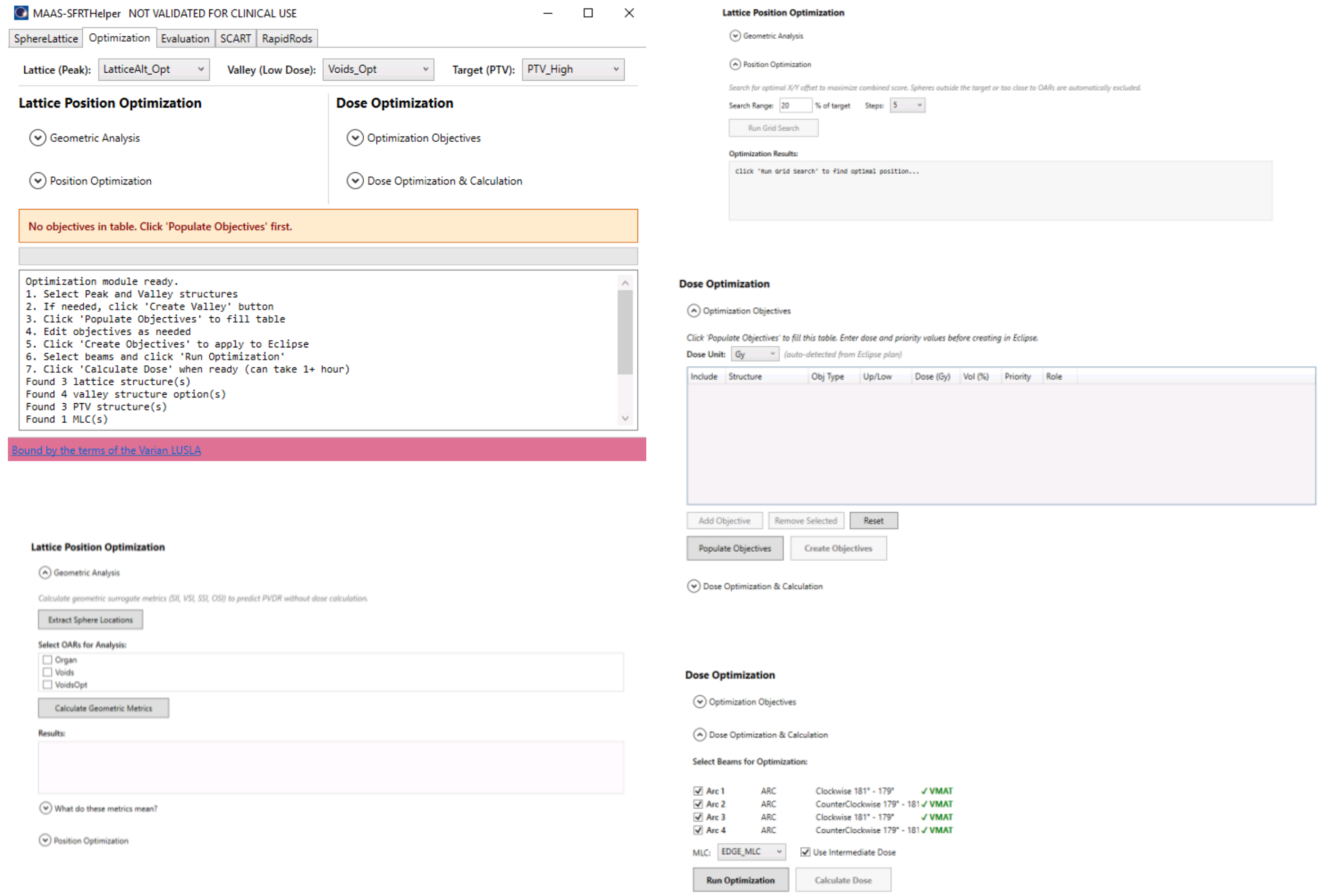

Figure 5: Optimization tab interface.

**Workflow.** The tab is divided into two panels as shown in Figure 5. Structure selection is shared across the top: the user selects a lattice structure, a valley, and a target PTV, all auto-populated from the SphereLattice tab when present. A valley structure is either inherited from the SphereLattice voids or created as the Boolean difference between the PTV and the lattice. The left panel handles lattice position optimization. The grid search is configured and run, returning candidate positions ranked by a composite geometric score. The user applies the preferred configuration, which rewrites the lattice at the chosen offset. The right panel handles dose optimization. An editable objectives table lists every structure in the set. Each structure is auto-assigned a role based on its name. Peak and Valley structures receive both point and mean objective rows. The user enters dose values and priorities directly. The dose unit is auto-detected from the Eclipse plan as Gy or cGy, with manual override available. VMAT optimization and dose calculation are then launched through ESAPI, producing a plan ready for the Evaluation tab (Section 2.4).

**Geometric surrogates.** The geometric search uses four surrogate metrics to score a candidate lattice configuration without requiring dose calculation. The underlying question is how the lattice looks from the beam's perspective: at each gantry angle, the beam projects three-dimensional sphere positions into two-dimensional shadows on the plane perpendicular to the beam direction. If two sphere shadows overlap from a given angle, the MLC cannot independently modulate dose to each sphere at that angle; if a sphere shadow falls on top of a nearby OAR, the beam must pass through a high-dose region to reach that organ. Three of the four metrics capture these effects. At gantry angle $\theta$, each sphere center at position $(x, y, z)$ projects to a two-dimensional coordinate $u = x\ cos\theta + y\ sin\theta; v = z$, where $u$ is the position across the beam and $v$ is the position along the treatment couch. The gantry rotates in the x-y plane; $v$ is invariant under this rotation. The sphere radius is preserved under the projection. These three metrics are computed every 5° around the arc (72 angles) and averaged.

Consider N spheres whose projections at a given angle produce circles with areas $A_i$ with $i = 1 \dots N$. Let $A_{ij}$ denote the intersection area of circles $i$ and $j$, computed analytically from their center-to-center distance and radii using the standard lens formula for two circles. The Sphere Isolation Index (SII) measures how much the projected circles overlap one another:

$$SII = 1 - \frac{\sum A_{ij}}{\sum A_i}. \quad (1)$$

An SII of 1.0 means no projected sphere overlaps any other at that angle, so each can be independently targeted by the MLC. An SII near 0 means most of the projected area is shared between sphere pairs.

The Valley Space Index (VSI) measures the fraction of the projected target that is not covered by any sphere projection. Let $A_T$ denote the projected area of the target (approximated as a circle with an equivalent radius). The union of all sphere projections is estimated as

$$A_U = \sum A_i - \sum A_{ij}. \quad (2)$$

and VSI is defined as

$$VSI = 1 - \frac{A_U}{A_T}. \quad (3)$$

Higher VSI means more of the target is visible as valley, giving the optimizer room to drive valley dose down.

The OAR Sparing Index (OSI) measures how much projected OAR is shadowed by sphere projections. Each OAR is first reduced to two scalar dimensions derived from its ESAPI bounding box: a lateral radius equal to the average of the X and Y extents divided by two, and a height equal to the Z extent. At each gantry angle, these two dimensions define an ellipse on the BEV plane with semi-axes equal to the lateral radius and half the height. To compute the sphere-OAR intersection area, the ellipse is approximated as a circle with geometric-mean radius, the intersection is computed using the standard circle-circle lens formula, and the result is scaled by an aspect-ratio correction. Projecting the actual OAR contours onto the BEV plane at each angle would capture irregular and concave shapes accurately but would require collecting all contour points in three dimensions, computing a projected boundary, and solving polygon-circle intersections at every candidate position, angle, and OAR. The bounding-box approach trades that accuracy for speed in a grid search that evaluates multiple configurations. The approximations are applied identically at every candidate position, preserving the relative ranking that drives the search. The total sphere-OAR intersection area $A_{SO}$ and the approximate projected OAR area $A_O$ give

$$OSI = 1 - \frac{A_{SO}}{A_O}. \tag{4}$$

An OSI of 1.0 means no sphere shadows the OAR at that angle; an OSI near 0 means the OAR is fully covered. When no OARs are selected, OSI defaults to 1.0. When multiple OARs are present, each receives its own OSI and the combined score is the average across all OARs.

The fourth metric, the Sphere Spread Index (SSI), is angle-independent and serves as a counterweight to the three BEV metrics. Because SII, VSI, and OSI can all improve when spheres are dropped from the lattice - fewer spheres imply less overlap, more open valley, and less OAR shadowing - a search guided by those metrics alone could converge on configurations with too few spheres to be clinically useful. SSI counteracts this: it yields a low score when spheres are lost or the lattice drifts off-center, pulling the composite score down even when the other three metrics improve. The volume component is

$$SSI_V = \min\left(1, \frac{\sum V_{Si}}{f_r V_T}\right), \tag{5}$$

where $V_{Si} = \frac{4}{3}\pi r_i^3$ is the volume of sphere $i$ and $V_T$ is the target volume; and $f_r$ is a user-adjustable reference fill ratio. The alignment component is

$$SSI_A = 1 - \frac{||C_L - C_T||}{R_T}, \tag{6}$$

where $C_L$ is the centroid of the sphere positions, $C_T$ and $R_T$ are the centroid and equivalent radius of the target, and the result is restricted to [0, 1]. The combined SSI is

$$SSI = \frac{1}{2}(SSI_V + SSI_A). \tag{7}$$

All four metrics lie in [0, 1] with higher values indicating more favorable geometry. They are combined into a single score

$$S = w_1 SII + w_2 VSI + w_3 OSI + w_4 SSI, \quad (8)$$

with default weights $w_1 = w_2 = w_3 = w_4 = 0.25$.

**Grid search.** The grid search translates the entire lattice in the x–y plane while keeping sphere radius and relative arrangement fixed. The user specifies a search range as a percentage of the target radius (default ±20%) and a step count. At each candidate offset, spheres that fall outside the target or overlap a selected OAR are dropped. The four surrogate metrics are evaluated on the surviving configuration. Positions are ranked by the combined score S, with ties broken by sphere count and then by closeness to the original position. Two options are presented: Apply Best Overall selects the highest-scoring configuration regardless of how many spheres survived, and Apply Baseline returns to the original unshifted position. The chosen configuration is written back to the structure set. Void structures from the SphereLattice tab (when present) are translated by the same offset and cropped according to the target boundary. Rigid translations preserve their interstitial placement relative to the lattice; therefore, void positions are not evaluated independently during the search.

**Dose optimization.** With the chosen lattice written to the structure set, VMAT optimization and dose calculation are invoked through ESAPI within the plugin. The user selects beams and whether to recalculate intermediate dose; the plugin then calls the ESAPI optimization and dose calculation methods using the plan's current settings. The resulting plan is stored in Eclipse and is available for analysis in the Evaluation tab or directly in Eclipse's plan evaluation workspace.

**Validation.** Optimization was validated on a simple sphere-in-cylinder phantom. A simple cubic lattice with r = 8 mm produced 22 spheres at the baseline position. The grid search was run at the default ±20% range, evaluating 81 candidate positions in under one second. The Best Overall configuration shifted the lattice to a position where the combined geometric score was maximized; 14 spheres survived after dropping those that fell outside the target or overlapped the OAR. VMAT optimization and dose calculation were run on both configurations using identical beam and objective settings. The repositioned configuration achieved a PVDR of 2.77 compared to 2.13 at baseline (30% improvement), with 8% higher peak mean dose, 17% lower valley mean dose, 22% lower OAR mean dose, and 26% lower OAR max dose, while using 36% fewer spheres. Figures 5 and 6 show the Optimization tab interface, the baseline and repositioned sphere configurations, and the resulting dose distributions. The result demonstrates that the geometric surrogates identify lattice positions that translate into improved dose-based plan quality; it also suggests that sphere positioning may be as important as sphere count for SFRT plan quality. The validation does not establish optimal search ranges or metric weights for clinical use; those will vary by disease site and institutional protocol.

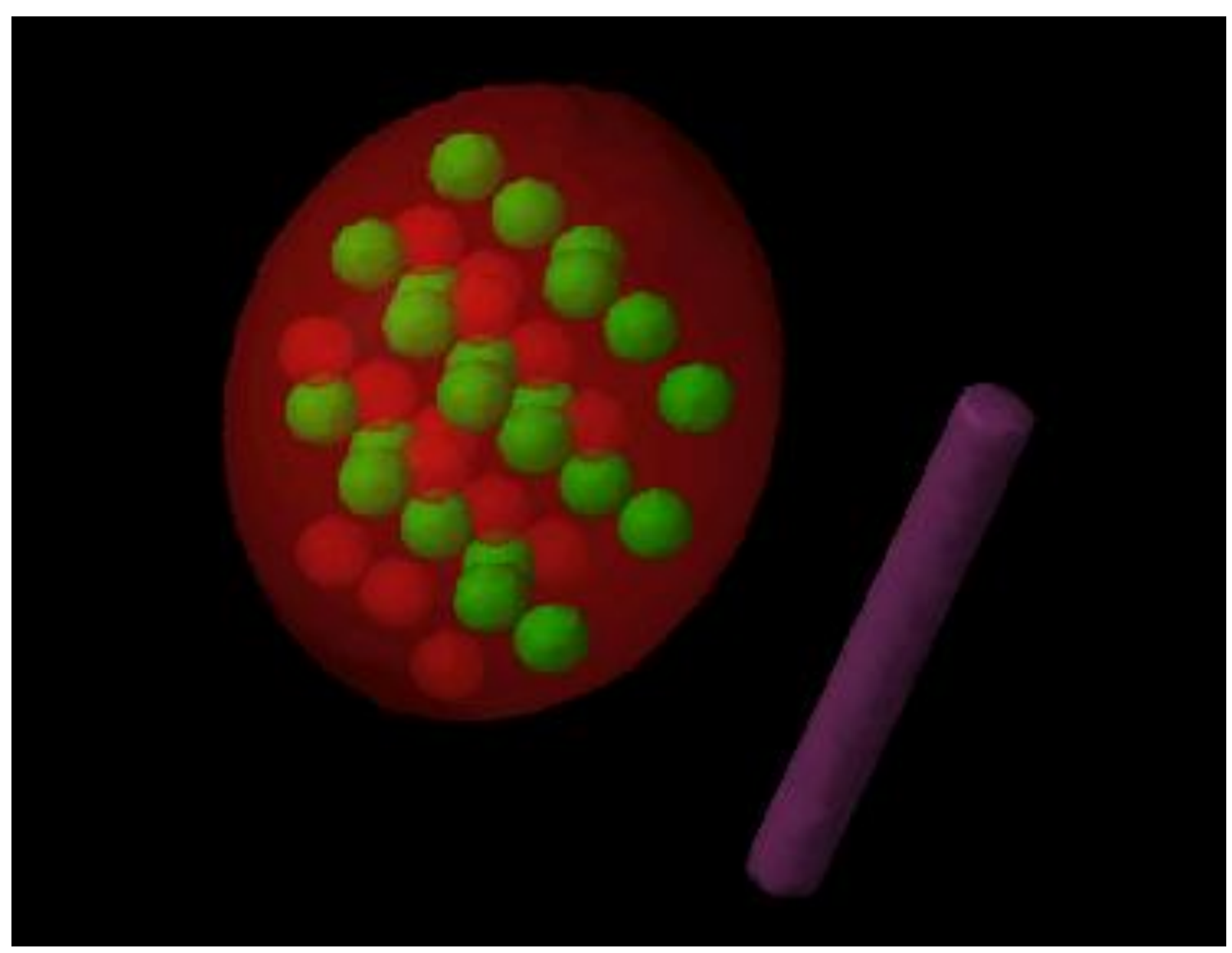

Figure 6: New lattice position following grid search – new are presented in red; old in green; OAR in magenta.

## 2.4 Evaluation

With the plan optimized and dose calculated, the final step is characterizing the resulting dose distribution. The peak-to-valley ratio reported for an SFRT plan depends on how peaks and valleys are defined. Some groups draw one-dimensional dose profiles through sphere centers and read peak and valley values off the profile [7, 14]; others threshold the three-dimensional dose distribution at a chosen level and call everything above a peak [1]; while others average dose inside peak and valley contours [27]. Each approach couples the reported ratio to the prescribed dose, the delivery technique, or the chosen threshold, making values difficult to compare across plans and institutions. The 2024 NRG Oncology/AAPM consensus identified this lack of standardized dosimetry reporting as a barrier to multi-institutional trials [5]. PVDR reported by the tab is directly comparable and reproducible across plans, dose levels, and delivery techniques.

**Analysis modes.** The tab provides four analysis modes that can be run on the same plan. 1D CAX plots dose along the central axis of a selected beam, producing a classical peak-valley profile through the lattice. 2D Multi-planar samples dose on a user-selected planar slice and displays peak-valley structure as a color map with interactive depth navigation. Dose Metrics computes scalar summary statistics over the full target volume [25]. The Evaluation also features 3D P/V, a geometrically grounded definition which defines a voxel as a peak if it lies inside a sphere in the lattice structure, and a valley otherwise. This method reports peak-valley metrics over the full target. 1D CAX and 2D Multi-planar require only the target structure and plan dose. Dose Metrics and 3D P/V additionally require the lattice structure; the novel 3D P/V analysis method obtains sphere centers and radius using the SphereExtractor service.

**Sphere extraction.** The service first queries ESAPI for the number of separate parts N in the lattice structure and the total structure volume V, then estimates a mean sphere radius as:

$$r = \sqrt[3]{\frac{3V}{4\pi N}}. \tag{9}$$

It then extracts per-slice contours of the lattice structure and computes the centroid and signed area of each contour using the shoelace formula [42]. Contours are clustered into spheres by grouping those whose XY centers fall within 0.5×$r$ and whose Z positions form a connected run with gaps no larger than 2.5×$r$. Each cluster's three-dimensional sphere center is the area-weighted centroid of its member contours - a calculation that yields sub-slice positional accuracy when the CT slice spacing is coarser than the sphere radius.

With sphere centers and radius, the tab samples the plan dose on a uniform three-dimensional grid spanning the target bounding box at a default resolution of 2 mm. Each grid point is first tested against the target boundary using the ESAPI IsPointInsideSegment method. Points outside the target are discarded. For each retained point, dose is sampled through the ESAPI GetDoseToPoint method and the point is classified geometrically: the squared Euclidean distance from the point to every sphere center is computed, and if the smallest such distance is less than or equal to $r^2$ the point is assigned to that sphere's peak bucket; otherwise it enters the valley pool. The result is a per-sphere list of peak voxels with their doses and a single valley pool from which the downstream metrics are computed. The classification is deterministic - repeated runs on the same target and plan return identical voxel assignments and identical metrics.

**3D peak-valley classification.** The metrics reported by 3D P/V follow from the per-sphere peak buckets and the valley pool. The effective peak-to-valley dose ratio (ePVDR) is a volume-weighted generalization of the traditional PVDR:

$$\text{ePVDR} = \frac{\sum_i \bar{D}_i V_i}{\bar{D}_{\text{valley}} \sum_i V_i}, \tag{10}$$

where $\bar{D}_i$ and $V_i$ are the mean dose and volume of the $i$-th sphere cluster and $\bar{D}_{\text{valley}}$ is the mean dose across the valley pool. A point PVDR is also reported as $D_{\text{peak,max}}/D_{\text{valley,min}}$, capturing the extreme peak-to-valley ratio rather than the average. Plan heterogeneity is quantified across all sampled voxels (peak and valley combined) by the heterogeneity index

$$\text{HI} = \frac{D_{\max} - D_{\min}}{D_{\max}}, \tag{11}$$

and the Coefficient of Variation (CV)

$$\text{CV} = \frac{\sigma}{\mu}, \tag{12}$$

with $\sigma$ and $\mu$ the standard deviation and mean of the sampled dose distribution. Mean peak separation - the average pairwise Euclidean distance between the N recovered sphere centers - provides a cross-check against the nominal lattice spacing. All metrics, together with per-sphere voxel counts, mean and max doses, and centroid coordinates, are exported to a CSV file.

**Onion-layer shell analysis.** An onion-layer shell analysis complements the peak-valley metrics with a radial view of dose heterogeneity inside the target. Each sampled voxel is assigned a Euclidean distance to the target center of mass. Voxels are sorted by distance and partitioned into five shells of equal voxel count. The tab reports mean dose, standard deviation, and coefficient of variation for each shell, and the user can step through shells interactively. This analysis requires only the plan dose and target structure.

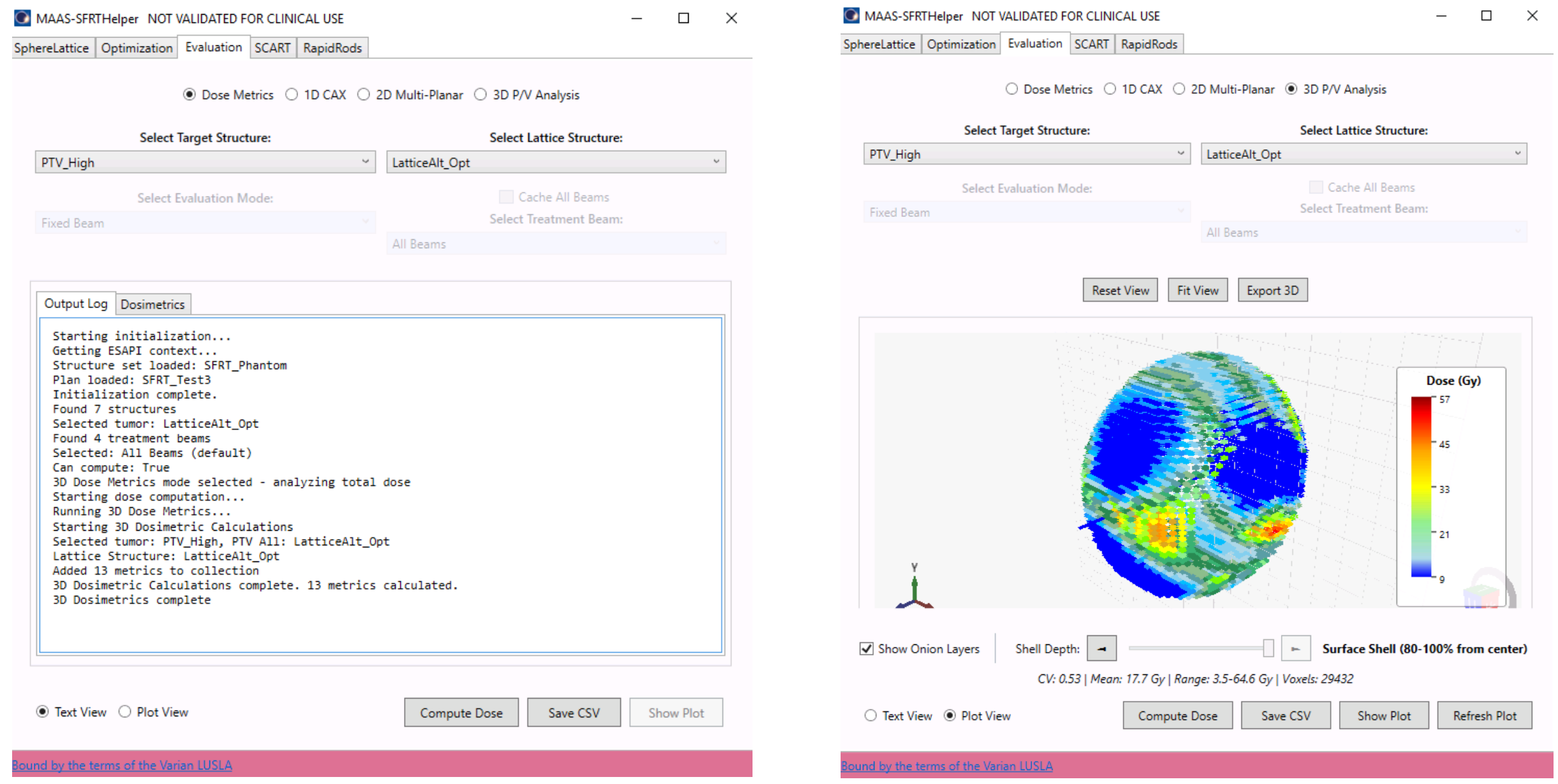

Figure 7: Evaluation tab interface.

| Metric | Ground-truth sphere count | Recalculated sphere count | Mean center position error [mm] | Max center position error [mm] |
|---|---|---|---|---|
| Value | 22 | 22 | 0.1 | 0.1 |

Table 2: Sphere extraction accuracy.

**Validation.** The Evaluation tab was validated on a synthetic phantom with a lattice generated by SphereLattice. Ground-truth sphere count, centers, and radii were exported via the SphereLattice CSV. We compared the recalculated sphere count against ground truth and measured center-position error in millimeters across all three axes. SphereExtractor identified all 22 spheres with a maximum center-position error of 0.1 mm, consistent with the area-weighted centroid converging within the CT slice spacing. The 3D peak/valley classification is deterministic by construction: the voxel grid, target boundary test, and nearest-sphere distance calculation involve no random sampling, so repeated runs on the same plan return identical voxel assignments and metrics. Validation of per-sphere dose agreement against Eclipse DVH readings is planned for a subsequent study with clinical data. Sphere count and position accuracy are reported in Table 2; the Evaluation tab interface is shown in Figure 7.

## 3. Data Format and Usage Notes

We distribute MAAS-SFRThelper as source code that the user compiles in Visual Studio against their local ESAPI installation. Building the code produces a single DLL with all third-party dependencies bundled. Alternatively, precompiled versions are available on the releases tab on GitHub. Script approval is required for the plugin to modify structure sets and invoke VMAT optimization and dose calculation. A structure set and plan must be loaded before launching the script. On first launch, the plugin directs the user to the project page, via hyperlink or QR code (for systems without internet access), to review the LUSLA and obtain an access code; entering the code in the plugin completes acceptance and enables use. Each release build carries an expiration date stamped at compile time; once that date passes, the plugin will not start until the user downloads a newer build from the GitHub repository. Source code is available at the GitHub URL in the Data Availability Statement.

Table 3 summarizes the software and hardware requirements. MAAS-SFRThelper depends only on Eclipse and the ESAPI runtime; no external services, databases, or network connections are required after installation. The hardware that Varian specifies for Eclipse itself is sufficient for the plugin.

| **Requirement** | **Specification** |
|---|---|
| Treatment planning system | Varian Eclipse 15.6 or higher |
| ESAPI license | Clinical or research scripting license |
| Operating system | Windows 10 or higher |
| .NET Framework | 4.6.1 or higher |
| Runtime dependencies | Prism, Helix Toolkit, NLog (bundled) |
| Installation | Compile source or prebuilt DLL |
| License | Varian LUSLA |

Table 3: System and software requirements for MAAS-SFRThelper.

The SphereLattice tab presents a dropdown of available structures from which the user selects the target and generation parameters. It writes the combined lattice, individual spheres, and void structures to the structure set through ESAPI to Eclipse. Export enables four CSVs to be saved to user's Documents or Desktop directory, covering generation parameters, vertex plus void positions in DICOM coordinates, per-sphere volumes, and summary statistics. SphereLattice disables generation when the sphere radius exceeds half the center-to-center spacing.

The Optimization tab includes position optimization, which requires a lattice structure, and a target PTV. It writes the repositioned lattice, valley, and translated void structures back to the structure set. This tab also includes dose optimization which requires at least one VMAT arc beam.

The Evaluation tab requires a plan with calculated dose. It exports a single CSV combining scalar metrics (ePVDR, HI, CV, mean peak separation) with per-cluster dose and position data.The Evaluation tab's 3D peak/valley mode requires that individual lattice spheres do not physically overlap, as sphere detection relies on identifying geometrically disconnected regions. In each case, the plugin halts the operation before any structures or plans are modified.

SCART requires a GTV with contours and creates its own beams from the plan's machine parameters. RapidRods requires a target structure and writes rod structures to the structure set. Each tab validates its inputs before running and reports problems as user-visible messages.

## 4. Discussion

MAAS-SFRThelper unifies three tasks of lattice SFRT planning inside a commercial treatment planning system: structure generation, position-aware optimization, and peak-valley evaluation. Five placement patterns accommodate variation in how institutions define and place lattices. The Evaluation tab defines peaks and valleys geometrically in addition to simple dose metrics. The novel PVDR method addresses the standardized-reporting gap identified by the 2024 NRG Oncology/AAPM consensus as a barrier to multi-institutional trials [5] by making plans comparable across prescriptions and delivery techniques. Geometric surrogate optimization brings position-aware planning into Eclipse, a capability previously demonstrated only in research codes [27,28]. The shared services architecture (Section 2.1) allows the Optimization and Evaluation tabs to reuse the same sphere-center recovery code, simplifying maintenance and future extension.

Several limitations should be noted. All validation was performed on digital phantoms. Preliminary patient analyses have been reported separately [34], but systematic clinical validation and multi-institutional testing remain important next steps. The plugin targets Eclipse only. Porting to other planning systems would require re-implementation against each vendor's scripting interface, or utilizing the dicom-based web port tool, LatticeRT [43]. The RapidRods tab and constraint-based Monte Carlo placement are experimental and should not be used clinically without site-specific validation. The geometrically grounded peak definition assumes lattice SFRT with discrete spherical vertices. It does not apply to GRID therapy with planar apertures or minibeam radiation therapy with narrow beamlets [1]. Sphere extraction relies on the ESAPI GetNumberOfSeparateParts method, which requires that spheres do not physically overlap.

Clinical validation with patient data is the most immediate need. Multiple patients with target volumes ranging from 233 to 8963 cc have been analyzed using the Evaluation tab [34], but per-sphere dose values have not yet been compared against Eclipse DVH values. Retrospective comparison against manually planned cases would provide the first clinical benchmark for the geometric surrogates and evaluation metrics. Integrating TG-263 nomenclature would allow the objectives table to auto-populate from standardized OAR names. Coupling the geometric surrogate search with influence-matrix optimization could close the gap to joint position-and-plan optimization [27,28]. Existing 1D CAX and 2D MultiPlanar evaluation modes are intended to extend the peak-valley evaluation framework to GRID therapy and minibeam radiation. However, these methods have not been extensively tested. Incorporating equivalent uniform dose (EUD) is a near-term goal for the evaluation tab. It could provide a single-number comparison alongside the geometric ePVDR for retrospective analysis and plan review. Correlating geometric surrogates with biological outcomes is a longer-term goal requiring multi-institutional data. The source code is available under the Varian LUSLA. Community contributions of new patterns, metrics, and validation datasets are encouraged.

## 5. Conclusion

MAAS-SFRThelper is a shared-source ESAPI plugin with several features, including: structure generation, position-aware optimization, and peak-valley evaluation into a unified workflow inside Eclipse. Five sphere placement patterns, a four-metric geometric surrogate search over candidate lattice positions, and a geometrically grounded three-dimensional peak-valley classification together support clinical SFRT planning that is consistent within an institution and transportable across institutions.

Validation on digital phantoms demonstrates that the structure generation, sphere extraction, geometric surrogate search, and evaluation metrics behave as specified. Shared-source distribution enables adoption by any site with an ESAPI-licensed Eclipse installation. As the clinical SFRT evidence base grows toward multi-institutional trials, standardized planning and reporting tools are a practical step toward the dosimetric consistency that the NRG Oncology/AAPM consensus has called for [5].

## Data Availability Statement

The MAAS-SFRThelper source code is publicly available at https://github.com/Varian-MedicalAffairsAppliedSolutions/MAAS-SFRThelper under the Varian Limited Use Software License Agreement (LUSLA). The LatticeRT source code and documentation are available at https://github.com/Varian-MedicalAffairsAppliedSolutions/LatticeRT.